\documentclass{aa}
\usepackage{natbib}
\usepackage{txfonts}
\usepackage{hyperref}
\usepackage{graphicx}
\graphicspath{{figures/}}

\newcommand*{\hd}{HD~43317\xspace}

\begin{document}

\title{Period spacings of gravity modes in rapidly rotating magnetic stars}
\subtitle{I.~Axisymmetric fossil field with poloidal and toroidal components}

\author{V. Prat\inst{1} \and S. Mathis\inst{1,2} \and B. Buysschaert\inst{2,3} \and J. Van Beeck\inst{3} \and D. M. Bowman\inst{3} \and C. Aerts\inst{3,4} \and C. Neiner\inst{2}}

\institute{
    AIM, CEA, CNRS, Universit\'e Paris-Saclay, Universit\'e Paris Diderot, Sorbonne Paris Cit\'e, F-91191 Gif-sur-Yvette, France
    \and
    LESIA, Observatoire de Paris, PSL University, CNRS, Sorbonne Universit\'e, Univ. Paris Diderot, Sorbonne Paris Cit\'e, 5 place Jules Janssen, F-92195 Meudon, France
    \and
    Instituut voor Sterrenkunde, KU Leuven, Celestijnenlaan 200D, 3001 Leuven, Belgium
    \and
    Dept. of Astrophysics, IMAPP, Radboud University Nijmegen, 6500 GL, Nijmegen, The Netherlands
}

\date{}

\abstract
{Stellar magnetic fields are  often invoked to explain the missing transport of angular momentum observed in  models of stellar interiors.
However, the properties of an internal magnetic field and the consequences of its presence on stellar evolution are largely unknown.}
{We study the effect of an axisymmetric internal magnetic field on the frequency of gravity modes in rapidly rotating stars to check whether gravity modes can be used to detect and probe such a field.}
{Rotation is taken into account using the traditional approximation of rotation and the effect of the magnetic field is computed using a perturbative approach.
As a proof of concept, we compute frequency shifts due to a mixed (i.e. with both poloidal and toroidal components) fossil magnetic field for a representative model of a known magnetic, rapidly rotating, slowly pulsating B-type star:  \hd.}
{We find that frequency shifts induced by the magnetic field scale with the square of its amplitude.
A magnetic field with a near-core strength of the  order of 150\,kG (which is consistent with the observed surface field strength of the order of 1\,kG) leads to signatures that are detectable in period spacings for high-radial-order gravity modes.}
{The predicted frequency shifts can be used to constrain internal magnetic fields and offer the potential for a significant step forward in our interpretation of the observed structure of gravity-mode period spacing patterns in rapidly rotating stars.}

\keywords{asteroseismology -- waves -- stars: magnetic field -- stars: oscillations -- stars: rotation}

\maketitle

\section{Introduction}

The development of helio- and asteroseismology in recent decades has drastically improved our knowledge of stellar interiors \citep[see e.g.][]{Aerts10, ChaplinMiglio, HekkerCD, Aerts19}.
In particular, seismic constraints on the internal rotation of red giants and subgiants \citep{Beck,Mosser,Deheuvels12, Deheuvels14, Deheuvels15, Triana17, Gehan} and intermediate-mass main-sequence stars \citep{Kurtz, Saio15, Murphy, VanReeth16, Aerts17, Ouazzani17} highlight a weak core-to-surface rotation contrast, which requires a much stronger transport of angular momentum  than is predicted by current models \citep[e.g.][]{Marques, Ceillier, Cantiello, Spada, Ouazzani18}.
For a few rapidly rotating, pulsating $\gamma$~Doradus ($\gamma$~Dor) stars, \citet{VanReeth18} also found very weak differential rotation, but the majority of the 37 analysed stars revealed uniform radial rotation profiles.
Two physical processes are good candidates to explain the missing transport of angular momentum: internal gravity waves \citep{LeeSaio93, Zahn97, TalonCharbonnel05, Pantillon, Mathis09, Lee, Rogers15} and stable or unstable magnetic fields \citep{Spruit, Heger, MathisZahn05, Fuller}.
We refer the reader to \citet{Aerts19} for a thorough review.

The improvements in the instrument sensitivity and the increasing number of stars included in spectropolarimetric campaigns allow us to detect magnetic fields at the surface of stars using the Zeeman effect \citep[see e.g.][]{DonatiLandstreet}.
For the massive O- and B-type stars, high-resolution spectropolarimetric surveys such as Magnetism in Massive Stars \citep[MiMeS;][]{Wade} and B fields in OB stars \citep[BOB;][]{Morel} have detected a large-scale magnetic field at the surface of approximately 7\% of the analysed stars.
Most of the detected fields around intermediate-mass and massive stars have a simple geometry, usually an inclined dipole, and a polar field strength between 300\,G and a few tens of kG.
These fields are probably  of fossil origin, since their properties do not scale with stellar parameters or rotation \citep{Mestel,Neiner15, EmeriauMathis}, in contrast with convective dynamo fields detected at the surface of low-mass stars.
However, a convective dynamo may still be present in the core \citep[e.g.][]{Brun05, Augustson16}.

Theoretical and numerical studies show that the stability of fossil magnetic fields requires that they extend deep within the stellar radiative envelope and  have a mixed configuration with both poloidal and toroidal components \citep{Tayler73, MarkeyTayler, Tayler80, BraithwaiteSpruit, Braithwaite07, Braithwaite09, DuezBM}.
The internal magnetic field of a massive star with a convective core and radiative envelope, however, is still largely unknown because stellar interiors are opaque to spectropolarimetric measurements.
Indirect information on the magnetic field can be obtained by constraining the transport of angular momentum and chemical species it may induce \citep[e.g.][]{Briquet12}.
Another promising way to constrain the structure of the internal magnetic field is to look at its effect on the oscillation modes of stars, in particular low-frequency gravity (\emph{g}) modes \citep[e.g.][]{Buysschaert18}.

In non-rotating stars, the internal magnetic field induces the splitting of \emph{g}-mode frequencies of the same angular degree but different azimuthal orders \citep{LedouxSimon}.
This has been applied  for a model representative of slowly pulsating B-type (SPB) stars with a purely poloidal dipolar field (e.g.  \citealt{Hasan}).

In slowly rotating, weakly magnetic stars, rotation and the magnetic field can both be treated as perturbations.
This has been mainly done for pressure (\emph{p}) modes in the Sun and rapidly oscillating Ap (roAp) stars, with either an axisymmetric toroidal field \citep{GoughTaylor} or an oblique (not aligned with the rotation axis) dipolar field \citep{DziembowskiGoode,GoughThompson,GoodeThompson,ShibahashiTakata}.
A similar treatment was applied to $\beta$ Cephei by \citet{ShibahashiAerts}.

When the full effects of rotation need to be taken into account, that is for rapidly rotating stars (such as $\gamma$~Dor, SPB, or Be stars) or very low-frequency modes (such as Rossby modes, also called \emph{r} modes), more complex formalisms have to be used.
\citet{Schenk} proposed such a formalism to describe the coupling between modes of rotating stars due to external forces.
\citet{Morsink} used this formalism to investigate the effect of a general, non-perturbative magnetic field on \emph{r} modes.

The rotationally perturbed \emph{g}-mode pulsation frequencies in rapidly rotating stars are better described using the traditional approximation of rotation (TAR) than a  perturbative approach.
This approximation assumes that the radial component of the Coriolis force is negligible with respect to the buoyancy force and that radial displacements are limited by buoyancy and are small compared to horizontal displacements \citep{Eckart, Townsend03}.
Thus, the horizontal component of the rotation vector is neglected.
The TAR also assumes that the star is spherically symmetric, which is reasonable only for moderately fast rotators because of centrifugal deformation.
This allows for separation of variables in spherical coordinates and more efficient computation of eigenmodes than when considering the full effects of rotation.
We refer the reader to \citet{Mathis08} and  \citet{Ouazzani17}, and references therein for a complete discussion of the TAR.

A standard tool used to interpret spectra of gravity modes is period spacing patterns, i.e. the morphology of period differences between modes of consecutive radial orders and the same angular degree and azimuthal order.
In chemically homogeneous, non-rotating stars, we expect to have a constant asymptotic period spacing, as derived from the asymptotic relations by \citet{Tassoul}.
However, in rotating stars, period spacing values seen in the inertial frame are not constant, and their slope as a function of the period is related to the rotation \citep{Bouabid, VanReeth15a, VanReeth15b, Ouazzani17}.
As an alternative approach to fitting the gradient of the period spacing pattern, \citet{Christophe} proposed a method of recovering uniform period spacings by stretching the pulsation periods.

Recently, \citet{Buysschaert18} considered the magnetic, rapidly rotating SPB star \hd, following the earlier studies of \citet{Papics12}, \citet{Briquet13}, and \citet{Buysschaert17}.
They computed magnetic splittings due to an axisymmetric, purely poloidal dipolar field in the non-rotating case \citep[following the perturbative formalism of][]{Hasan} and found that they were negligible with respect to rotational splittings, thus justifying the perturbative approach.
The authors combined photometric and spectroscopic time series to perform mode identification and modelled the star using a grid of non-rotating, non-magnetic equilibrium stellar structure models.
They found tentative evidence for a low amount of convective core overshooting, which was interpreted as being caused by the large-scale magnetic field in \hd \citep[see also][]{Briquet12}.
However, as already mentioned above, such a purely poloidal magnetic field would be unstable.
In addition, rapid rotation most likely influences the effect of the magnetic field on stellar oscillations.
The goal of the present study is to develop a perturbative description of gravito-inertial modes (i.e. gravity modes affected by rotation) in the presence of a stable, mixed (i.e. with both poloidal and toroidal components), axisymmetric magnetic field within the TAR.

We present how the perturbation theory is modified in the presence of rotation in Sect.~\ref{sec:perturb}.
We show how we derived the expression of the frequency shifts induced by an axisymmetric magnetic field in Sect.~\ref{sec:split}.
Then, we apply our new formalism to a representative stellar structure model of \hd found by \citet{Buysschaert18} in Sect.~\ref{sec:appl}.
We discuss the results and prospects in Sect.~\ref{sec:conc}.

\section{First-order perturbation theory of rotating stars}
\label{sec:perturb}

In the uniformly rotating case, assuming a Lagrangian displacement of the form $\vec\xi=\vec{\hat\xi} e^{-i\omega t}$, where $\omega$ is the angular frequency in the corotating frame and $t$ the time, the linearised momentum equation can be written as
\begin{equation}
    \label{eq:eigen_rot}
    \omega^2\vec\xi+i\omega\vec{\mathcal{B}}(\vec\xi)+\vec{\mathcal{C}}(\vec\xi)=0,
\end{equation}
where $\vec{\mathcal{B}}(\vec\xi)=2\vec\Omega\wedge\vec\xi$ is the Coriolis operator, $\vec\Omega=\Omega\vec e_z$ is the rotation vector, and $\vec{\mathcal{C}}(\vec\xi)$ describes forces that do not depend on the frequency.
In the non-magnetic case it reduces to pressure and buoyancy forces, $\vec{\mathcal{C}}_0(\vec\xi)=-\vec\nabla\delta P/\rho+\delta\rho\vec\nabla P/\rho^2$, where $\delta P$ and $\delta\rho$ are respectively the pressure and density Eulerian perturbations around the equilibrium values $P$ and $\rho$.
In the TAR, the horizontal component of the rotation vector is neglected, so that the Coriolis operator reduces to $\vec{\mathcal{B}}(\vec\xi)=2\Omega\cos\theta\vec e_r\wedge\vec\xi$, where $\theta$ is the colatitude and $\vec e_r$ is the radial unit vector.

The first-order perturbation by an additional force that does not depend on $\omega$ is obtained by writing $\omega=\omega_0+\varepsilon\omega_1$, $\vec\xi=\vec\xi_0+\varepsilon\vec\xi_1$, and $\vec{\mathcal{C}}=\vec{\mathcal{C}}_0+\varepsilon\vec{\mathcal{C}}_1$, where $\varepsilon\ll1$.
At the zeroth order in $\varepsilon$, Eq.~\eqref{eq:eigen_rot} leads to the unperturbed equation
\begin{equation}
    \label{eq:eigen_rot0}
    \omega_0^2\vec\xi_0+i\omega_0\vec{\mathcal{B}}(\vec\xi_0)+\vec{\mathcal{C}}_0(\vec\xi_0)=0,
\end{equation}
and the first-order equation reads
\begin{equation}
-\omega_1[2\omega_0\vec\xi_0+i\vec{\mathcal{B}}(\vec\xi_0)]=\vec{\mathcal{C}}_1(\vec\xi_0)+\omega_0^2\vec\xi_1+i\omega_0\vec{\mathcal{B}}(\vec\xi_1)+\vec{\mathcal{C}}_0(\vec\xi_1).
\end{equation}
Given that $i\vec{\mathcal{B}}$ and $\vec{\mathcal{C}}_0$ are Hermitian, Eq.~\eqref{eq:eigen_rot0} allows us to simplify the scalar product of the previous equation with $\vec\xi_0$, which leads to
\begin{equation}
    \label{eq:gen_split}
    \omega_1 = -\frac{\langle\vec\xi_0,\vec{\mathcal{C}}_1(\vec\xi_0)\rangle}{2\omega_0\langle\vec\xi_0,\vec\xi_0\rangle+\langle\vec\xi_0,i\vec{\mathcal{B}}(\vec\xi_0)\rangle},
\end{equation}
where the scalar product is defined by
\begin{equation}
    \langle\vec\xi,\vec\zeta\rangle=\int_V\rho\vec\xi^*\cdot\vec\zeta{\rm d}V,
\end{equation}
and the asterisk ($^*$) denotes the complex conjugate.

\section{Magnetic frequency shifts}
\label{sec:split}

In the TAR, unperturbed eigenmodes for gravito-inertial waves are given in spherical coordinates $(r,\theta,\varphi)$ by
\begin{equation}
    \vec\xi_0 = [\xi_{\rm r}(r) H_r(\theta), \xi_{\rm h}(r) H_\theta(\theta), i\xi_{\rm h}(r) H_\varphi(\theta)]e^{i(m\varphi-\omega_0 t)},
\end{equation}
where $m$ is the azimuthal order, and $H_r$, $H_\theta$, and $H_\varphi$ are radial, latitudinal, and azimuthal Hough functions, respectively \citep{Hough,LeeSaio97, Townsend03}.
Their definitions are given in Appendix~\ref{sec:hough}.

In the present work we consider a magnetic field that is weak enough such that the effect of the unperturbed Lorentz force on the equilibrium state is negligible \citep[e.g.][]{DuezMTC}.
In addition, for simplicity we assume for our analytical calculations that the field is axisymmetric and dipolar:
\begin{equation}
    \vec B = B_0[b_r(r)\cos\theta, b_\theta(r)\sin\theta, b_\varphi(r)\sin\theta].
\end{equation}
For our numerical calculations (see Sect.~\ref{sec:appl}), we use a field defined by
\begin{equation}
    \label{eq:field}
    \vec B = \frac{1}{r\sin\theta}\left(\vec\nabla\psi\wedge\vec e_\varphi+\lambda\frac{\psi}{R}\vec e_\varphi\right),
\end{equation}
with $\psi$ being the stream function, which satisfies
\begin{equation}
    \label{eq:psi}
    \begin{aligned}
        \psi=-\mu_0\alpha\lambda\frac{r}{R}&\left[j_1\left(\lambda\frac{r}{R}\right)\int_r^Ry_1\left(\lambda\frac{x}{R}\right)\rho x^3{\rm d}x\right.\\
        &\quad\left.+y_1\left(\lambda\frac{r}{R}\right)\int_0^rj_1\left(\lambda\frac{x}{R}\right)\rho x^3{\rm d}x\right]\sin^2\theta,
    \end{aligned}
\end{equation}
where $\vec e_\varphi$ is the unit vector in the azimuthal direction, $R$ is the stellar radius, $\lambda\simeq35.89$ is the smallest non-negative constant such that $\vec B$ vanishes at the stellar surface, $\mu_0$ is the vacuum permeability, $\alpha$ is an amplitude scaling factor, and $j_1$ (respectively $y_1$) is the first-order spherical Bessel function of the first (respectively second) kind.
Defining $A$ such that $\psi=A(r)\sin^2\theta$, we can write $b_r = 2A/r^2$, $b_\theta=-A'/r$, and $b_\varphi=\lambda A/(rR)$,
where the prime symbol ($'$) denotes a total radial derivative.
One of the properties of this field is that the specific perturbed Lorentz force $\delta\vec F_{\rm L}/\rho$ (see Eq.~\eqref{eq:lorentz}) remains finite at the surface even though $\rho$ becomes very small, so the perturbative approach can be used {a priori}.

This field aims to represent a fossil field in the stellar radiative zone, although its origin is not explicitly used further.
Since purely poloidal and toroidal fields are known to be unstable \citep{Tayler73, MarkeyTayler}, we use the mixed configuration proposed by \citet{DuezBM}, which is based on energy\footnote{\citet{BroderickNarayan} and \citet{DuezMathis} have demonstrated that dipolar fields should be favoured for fossil fields, as shown by the observations, because they are the equilibrium states with the lowest energies.} and stability arguments.
This configuration has been demonstrated to be stable, using three-dimensional  simulations \citep{DuezBM}, and to have properties similar to those obtained using numerical simulations of the formation of fossil fields \citep{BraithwaiteSpruit, BraithwaiteNordlund, Braithwaite08}.
Although this configuration has been derived in the non-rotating case, numerical simulations and theoretical calculations show that similar magnetic equilibria can be obtained in the rotating case \citep{Duez, EmeriauMathis} even if their formation time can be modified \citep{BraithwaiteCantiello}.
Generally, these equilibria are not axisymmetric, but rather oblique dipoles.
However, as a first step, we restrict our study to an axisymmetric configuration, which is mathematically simpler.
Here the field extends down to the centre, but in reality the field would be more complex due to its interaction with the dynamo field likely present in the convective core \citep{Featherstone}.
For magnetic F- and A-type stars, the fossil field might also interact with the dynamo in the sub-surface convective envelope \citep{Augustson13}.

The induction equation in the ideal magneto-hydrodynamical approximation implies that the Eulerian perturbation to the magnetic field due to the oscillation displacement is
\begin{equation}
    \delta\vec B = \vec\nabla\wedge(\vec\xi_0 \wedge \vec B).
\end{equation}
The perturbed Lorentz force reads
\begin{equation}
    \label{eq:lorentz}
    \delta\vec F_{\rm L} = \frac{1}{\mu_0}[(\vec\nabla\wedge\vec B)\wedge\delta\vec B + (\vec\nabla\wedge\delta\vec B)\wedge\vec B].
\end{equation}
Equation~\eqref{eq:gen_split} implies that magnetic frequency shifts $\delta\omega=\varepsilon\omega_1$ are proportional to $\langle\vec\xi_0,\delta\vec F_{\rm L}/\rho\rangle=\langle\vec\xi_0,\varepsilon\vec{\mathcal{C}}_1(\vec\xi_0)\rangle$.
Therefore, the frequency shifts are proportional to the square of the magnetic field amplitude.

The computation of the magnetic frequency shifts involves a large number of terms.
A significant fraction of them are zero for symmetry reasons.
\citet{Hasan} considered only one term, which they claimed was dominant for high-radial-order \emph{g} modes, i.e. for modes with an absolute value of the radial order\footnote{By convention, \emph{g} modes have negative radial orders, in contrast with \emph{p} modes, which have positive ones.} $|n|$ much larger than the angular degree $\ell$.
Since the effect of rotation may change this, we choose to keep all non-zero terms.
They are listed in Appendix~\ref{sec:terms}.

The first term of the denominator in Eq.~\eqref{eq:gen_split} involves the scalar product
\begin{equation}
    \langle\vec\xi_0,\vec\xi_0\rangle = 2\pi\int_0^R\int_0^\pi\rho r^2\left[|\xi_{\rm r}|^2 H_r^2 + |\xi_{\rm h}|^2 (H_\theta^2+H_\varphi^2)\right]{\rm d}r\sin\theta{\rm d}\theta.
\end{equation}
For high-radial-order \emph{g} modes, the first (radial) term is much smaller than the second (horizontal) one, and can thus be neglected, as done in \citet{Hasan}.
For low-radial-order modes (i.e. $|n|<5$), though, this introduces significant errors.
Note that in the model for \hd \citep{Buysschaert18}, some modes had such low radial orders.

The second term of the denominator in Eq.~\eqref{eq:gen_split} yields
\begin{equation}
    \langle\vec\xi_0,i\vec{\mathcal{B}}(\vec\xi_0)\rangle=8\pi\Omega\int_0^R\rho r^2|\xi_{\rm h}|^2{\rm d}r\int_0^\pi H_\theta H_\varphi\sin\theta\cos\theta{\rm d}\theta.
\end{equation}
Since this term scales with the spin factor $2\Omega/\omega$, it is negligible for low-radial-order modes, but not for high-radial-order ones.
When it is negligible, Eq.~\eqref{eq:gen_split} implies
\begin{equation}
    \label{eq:scaling}
    \frac{\delta\omega}{\omega_0}\propto \frac{B_0^2}{\omega_0^2},
\end{equation}
as in the non-rotating case, where \citet{Hasan} found that
\begin{equation}
    \label{eq:split_coeff}
    \frac{\delta\omega}{\omega_0}=S_{\rm c}B_0^2.
\end{equation}
This defines the splitting coefficient $S_{\rm c}$, which is proportional to $\mathcal{I}/\omega_0^2$ in their study, with
\begin{equation}
    \label{eq:int_ratio}
    \mathcal{I} = \frac{\int_0^R|(rb_r\xi_{\rm h})'|^2{\rm d}r}{\int_0^R|\xi_{\rm h}|^2(\rho/\rho_{\rm c})r^2{\rm d}r},
\end{equation}
where $\rho_{\rm c}$ is the central density of the star.
It follows that the effect of the  magnetic field is stronger at lower frequencies, i.e. longer periods or higher radial orders.
This is consistent with the fact that those frequencies are closer to the Alfv\'en frequency, which characterises the propagation of magnetic waves.

High-radial-order modes have a large radial wavenumber, which allows us to perform a Jeffreys-Wentzel-Kramers-Brillouin small-wavelength analysis \citep[see e.g.][in the context of stellar oscillations]{Unno}.
When the poloidal component of the magnetic field is much larger than the toroidal component or of the same order of magnitude, which is the case here, the dominant term of the numerator in Eq.~\eqref{eq:gen_split} is proportional to
\begin{equation}
    \int_0^R|(rb_r\xi_{\rm h})'|^2{\rm d}r\int_0^\pi(H_\theta^2+H_\varphi^2)\cos^2\theta\sin\theta{\rm d}\theta.
\end{equation}
This is consistent with the result obtained in the non-rotating case by \citet{Hasan}.
However, for sub-inertial ($\omega<2\Omega$) gravito-inertial waves, which are often  excited in rapidly rotating stars \citep[e.g.][]{Neiner12,Moravveji16,Saio18}, Hough functions differ significantly from spherical harmonics and our more general formalism is needed.
These waves are trapped in an equatorial belt.
In contrast, when the toroidal component is much larger than the poloidal component, the analysis predicts that eight other terms of the numerator in Eq.~\eqref{eq:gen_split} could have a significant impact on the perturbation (see Appendix~\ref{sec:tor}).

\section{Application to \hd: proof of concept}
\label{sec:appl}

As a proof of concept, we now apply our new formalism to a representative stellar model of \hd, based on the analysis by \citet{Buysschaert18}, which is a rapidly rotating, magnetic B-type star exhibiting \emph{g} modes.
The photometric and spectroscopic analysis of this star by \citet{Papics12} indicated a solar-like metallicity $Z=0.014$, an effective temperature $T_{\rm eff}=17350\pm750\,{\rm K}$, a surface gravity $\log g=4.0\pm0.1\,{\rm dex}$, and a rotation period $P_{\rm rot}=0.897673(4)\,{\rm d}$ (about 60\% of the Roche critical rotation rate).
The spectropolarimetric analysis \citep{Briquet13,Buysschaert17} showed the presence of a dipolar surface magnetic field of $1312\pm332\,{\rm G}$, and \citet{Buysschaert18} estimated the obliquity angle as $\beta=81\pm6\,^\circ$.
The best model computed by \citet{Buysschaert18} has a stellar mass $M_\star=5.8\,{\rm M}_\odot$, a central hydrogen mass fraction $X_{\rm c}=0.54$, and an exponential convective core overshooting parameter $f_{\rm ov}=0.004$.
The stellar model was evaluated using a grid-based approach with the Modules for Experiments in Stellar Astrophysics (MESA) one-dimensional  stellar structure and evolution code \citep{Paxton18}, and the eigenmodes were computed with the GYRE oscillation code \citep{TownsendTeitler, Townsend18}.

\begin{figure}
    \resizebox{\hsize}{!}{\includegraphics{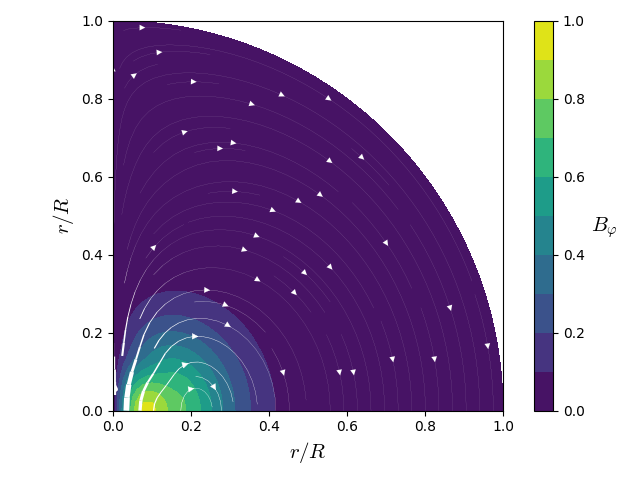}}
    \caption{Representation of the magnetic field used in the numerical computations.
        The stream lines represent the poloidal component of the field (their line width is proportional to the amplitude).
        The background coloured contours represent the toroidal component.
        In this plot, the field is normalised so that the maximum of its norm is unity.
        Both poloidal and toroidal components have a maximum amplitude of order unity.
        This field is scaled using the amplitude parameter $\alpha$ mentioned in Eq.~\eqref{eq:psi}.}
    \label{fig:mag}
\end{figure}

The magnetic field obtained from Eq.~\eqref{eq:field} with the density profile of the MESA model is illustrated in Fig.~\ref{fig:mag}.
We note that the obliquity of the field with respect to the rotation axis is not accounted for in the present study.
In the following, unless mentioned otherwise, we fix the maximum value for the internal magnetic field to $B_{\rm max}=1.5\cdot10^5\,{\rm G}$, which is approximately 100 times the value of the surface magnetic field detected by spectropolarimetry.
This  ratio is consistent with numerical simulations of magnetic equilibria \citep{Braithwaite08}, but our value of $B_0$ ($\simeq B_{\rm max}$) is significantly larger than the values (26.1\,kG and 82.4\,kG) used to compute magnetic splittings in \citet{Buysschaert18}.

We numerically computed the frequency perturbations of dipole ($\ell=1$) and quadrupole ($\ell=2$) modes.
For non-axisymmetric modes, it is important to distinguish between the frequency in the corotating frame $\omega$ and the frequency in the inertial frame $\omega_{\rm i}$, where
\begin{equation}
    \label{eq:doppler}
    \omega_{\rm i} = \omega + m\Omega.
\end{equation}
\citet{Buysschaert18} incorrectly computed the magnetic frequency shifts of the identified \emph{g} modes of \hd.
A confusion concerning the numbering of GYRE output files led the authors to use incorrect eigenmodes in their computations.
Our Table~\ref{tab:obs}  gives a corrected version of their Table~3, which compares rotational splittings with magnetic splittings in the non-rotating case.
We also add the magnetic frequency shifts in the TAR computed in the present paper.
\begin{table*}[t]
\caption{Comparison of rotational frequency shifts, magnetic frequency shifts in the non-rotating case, and magnetic frequency shifts in presence of rotation.
    For each identified frequency in the spectrum of \hd are given the radial order $n$, the angular degree $\ell$, the azimuthal order $m$, the theoretical frequency computed with GYRE, the corresponding frequency at zero rotation, the rotational frequency shift, magnetic frequency shifts computed in the non-rotating case with two different values of the magnetic field, the magnetic splitting coefficient $S_{\rm c}$ defined in Eq.~\eqref{eq:split_coeff}, the quantity $\mathcal{I}$ defined in Eq.~\eqref{eq:int_ratio}, and the magnetic frequency shift in the TAR computed using the formalism developed in the present paper.
    Frequencies are given in ${\rm d}^{-1}$.}
\centering
\(
\begin{array}{rlrl|ll|llcc|l}
\hline
\hline
n & \ell & m & f_{n, \ell, m}  & f_{n, \ell} & f_{\rm shift}\tablefootmark{a} & f_{\rm shift}\tablefootmark{bc} & f_{\rm shift}\tablefootmark{bc} & S_{\rm c}\tablefootmark{bc} & \mathcal{I}\tablefootmark{b}  &  f_{\rm shift} \\
&      &   & \text{rot.}         & \text{no rot.}      & \text{rot.}   & 26.1\,{\rm kG}    & 82.4\,{\rm kG} &   [{\rm G}^{-2}]  & & \text{mag./rot.}    \\
\hline
-11     &       1       &       -1      &       0.6867  &       0.8187  &       0.1320  &       5.2442  \cdot10^{-5} &       5.2270 \cdot10^{-4} &   9.4030  \cdot10^{-14}   &       1.9334  \cdot10^{4} &   1.0268 \cdot10^{-5}  \\
-10     &       1       &       -1      &       0.7573  &       0.8979  &       0.1406  &       5.6099  \cdot10^{-5} &       5.5915 \cdot10^{-4} &   9.1720  \cdot10^{-14}   &       2.2684  \cdot10^{4}     &   1.1100 \cdot10^{-5} \\
-9      &       1       &       -1      &       0.8381  &       0.9923  &       0.1542  &       9.6437  \cdot10^{-5} &       9.6120 \cdot10^{-4} &   1.4267  \cdot10^{-13}   &       4.3094  \cdot10^{4}     &   1.7750 \cdot10^{-5} \\
-15     &       2       &       -1      &       0.8720  &       1.0462  &       0.1742  &       4.4627  \cdot10^{-4} &       4.4481 \cdot10^{-3} &   6.2618  \cdot10^{-13}   &       2.2076  \cdot10^{5}     &   2.9489 \cdot10^{-4} \\
-8      &       1       &       -1      &       0.9222  &       1.0933  &       0.1712  &       2.1844  \cdot10^{-4} & 2.1772 \cdot10^{-3} &   2.9329  \cdot10^{-13}   &       1.0754  \cdot10^{5}     &   4.8303 \cdot10^{-5} \\
-7      &       1       &       -1      &       1.0037  &       1.1943  &       0.1905  &       1.2655  \cdot10^{-4} &       1.2613 \cdot10^{-3} &   1.5555  \cdot10^{-13}   &       6.8059  \cdot10^{4}     &   5.3927 \cdot10^{-5} \\
-11     &       2       &       -1      &       1.1268  &       1.4108  &       0.2839  &       2.8547  \cdot10^{-5} & 2.8453 \cdot10^{-4} &   2.9705  \cdot10^{-14}   &       1.9044  \cdot10^{4}     &   1.1579 \cdot10^{-5} \\
-6      &       1       &       -1      &       1.1483  &       1.3708  &       0.2225  &       2.5492  \cdot10^{-5} &       2.5408 \cdot10^{-4} &   2.7300  \cdot10^{-14}   &       1.5736  \cdot10^{4}     &   1.0588 \cdot10^{-5} \\
-10     &       2       &       -1      &       1.2198  &       1.5454  &       0.3256  &       3.0370  \cdot10^{-5} &       3.0270 \cdot10^{-4} &   2.8847  \cdot10^{-14}   &       2.2191  \cdot10^{4}     &   1.1646 \cdot10^{-5} \\
-9      &       2       &       -1      &       1.3337  &       1.7058  &       0.3721  &       5.1999  \cdot10^{-5} &       5.1829 \cdot10^{-4} &   4.4749  \cdot10^{-14}   &       4.1940  \cdot10^{4}     &   1.7241 \cdot10^{-5} \\
-5      &       1       &       -1      &       1.3775  &       1.6455  &       0.2680  &       8.9291  \cdot10^{-6} &       8.8998 \cdot10^{-5} &   7.9658  \cdot10^{-15}   &       6.6164  \cdot10^{3}     &   2.5934 \cdot10^{-6} \\
-4      &       1       &       -1      &       1.7358  &       2.0601  &       0.3244  &       4.4448  \cdot10^{-6} &       4.4302 \cdot10^{-5} &   3.1672  \cdot10^{-15}   &       4.1234  \cdot10^{3}     &   1.0487 \cdot10^{-6} \\
-6      &       2       &       -1      &       1.8191  &       2.3408  &       0.5217  &       1.2792  \cdot10^{-5} &       1.2750 \cdot10^{-4} &   8.0223  \cdot10^{-15}   &       1.4158  \cdot10^{4}     &   8.9695 \cdot10^{-6} \\
-2      &       1       &       -1      &       3.4811  &       3.9638  &       0.4826  &       5.1817  \cdot10^{-6} &       5.1647 \cdot10^{-5} &   1.9191  \cdot10^{-15}   &       9.2493  \cdot10^{3}     &   3.7177 \cdot10^{-7} \\
-6      &       2       &       2       &       4.3408  &       2.3408  &       2.0000  &       7.9952  \cdot10^{-6} &       7.9690 \cdot10^{-5} &   5.0140  \cdot10^{-15}   &       1.4158  \cdot10^{4}     &   3.1559 \cdot10^{-6} \\
-1      &       1       &       -1      &       4.9948  &       6.1084  &       1.1135  &       1.7332  \cdot10^{-4} &       1.7275 \cdot10^{-3} &   4.1653  \cdot10^{-14}   &       4.7676  \cdot10^{5}     &   1.4840 \cdot10^{-7} \\
\hline
\end{array}
\)
\label{tab:obs}
\tablefoot{
    \tablefoottext{a}{In contrast with \citet{Buysschaert18}, where rotational frequency shifts were computed for the corresponding zonal ($m=0$) mode, we simply compute them here as the difference between the frequency in the rotating case and the frequency in the non-rotating case.}
    \tablefoottext{b}{These quantities were incorrectly computed in \citet{Buysschaert18} (see text).}
    \tablefoottext{c}{In \citet{Hasan}, the same notation $S_{\rm c}$ is used indiscriminately for the coefficient used to compute magnetic frequency shifts (in Sect.~2) and for the coefficient used to compute magnetic splittings between modes of different azimuthal orders (in Sect.~3).
Here we use the former definition, and the magnetic frequency shifts are computed accordingly.}
}
\end{table*}
Although the magnetic frequency shifts in the TAR sometimes show significant differences from those computed in the non-rotating case,  most of the time they are of the same order of magnitude, and remain small compared to the rotational shifts, which is consistent with our perturbative approach.

More generally, to check whether the perturbative approach is relevant, we compare the Alfv\'en frequency and the pulsation frequency $\omega$.
The Alfv\'en frequency is given by
\begin{equation}
    \omega_{\rm A}=\frac{\vec B\cdot\vec k}{\sqrt{\mu_0\rho}},
\end{equation}
where $\vec k$ is the wave vector, and must be much smaller than the pulsation frequency in the corotating frame.
For high-radial-order modes, $\vec B\cdot\vec k\simeq B_r k_r=b_r(r)k_r\cos\theta$.
Therefore, the magnitude of the Alfv\'en frequency is highest along the rotation axis, but high-radial-order modes are sub-inertial, and thus trapped in the region where $|\cos\theta| < \omega/(2\Omega)$ \citep[see e.g.][]{PratLB}.
The criterion reads
\begin{equation}
    \frac{b_rk_r}{2\Omega\sqrt{\mu_0\rho}} \ll 1.
\end{equation}
To go further, we use the rough estimate $k_r\sim|n|/R$.
The quantity $b_r/\sqrt{\rho}$ is highest at the centre of the star, but gravity waves are mostly sensitive to the near-core region.
In this region, $b_r/\sqrt{\rho}\simeq0.34B_0/\sqrt{\rho_{\rm c}}$, which finally leads to
\begin{equation}
    \label{eq:criterion}
    B_0 \ll \frac{\Omega R\sqrt{\mu_0\rho_{\rm c}}}{0.17|n|}.
\end{equation}
This criterion is verified for all the modes computed in this work, which justifies the perturbative treatment {a priori}.

In the following sections we study more deeply the magnetic frequency shifts of zonal (Sect.~\ref{sec:zonal}), prograde (Sect.~\ref{sec:pro}), and retrograde modes (Sect.~\ref{sec:retro}).
We investigate the effect of rotation in Sect.~\ref{sec:rot}.

\subsection{Zonal modes}
\label{sec:zonal}

\begin{figure}
    \resizebox{\hsize}{!}{\includegraphics{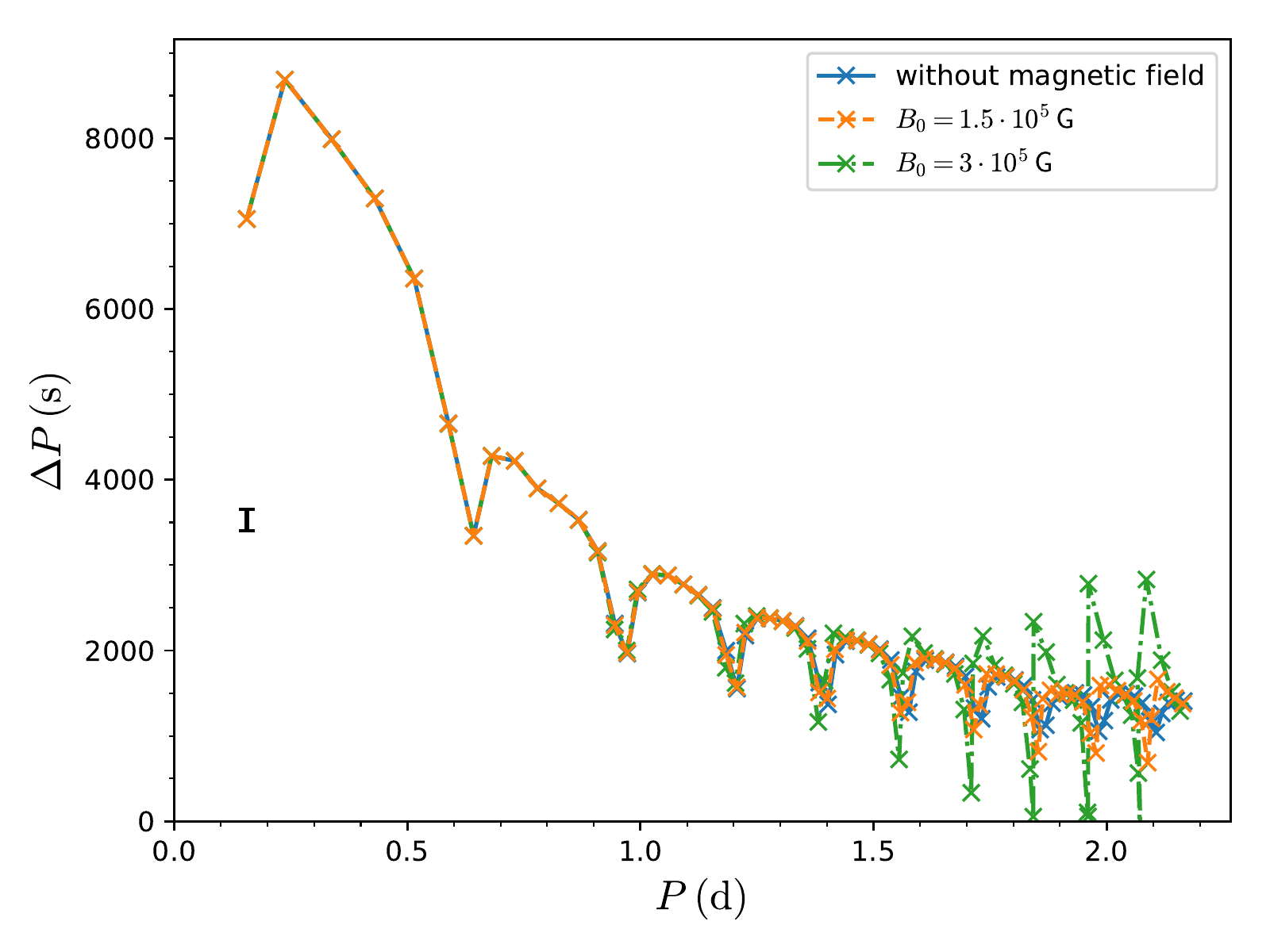}}
    \caption{Period spacings of \emph{g} modes with radial orders from -1 (left) to -74 (right) as a function of the period in the inertial frame for $\ell=1$, $m=0$, and different magnetic field strengths.
    The vertical black bar on the left represents a typical observational error bar of 250\,s \citep{VanReeth15b}.}
    \label{fig:l1m0}
\end{figure}

Figure~\ref{fig:l1m0} represents for $\ell=1$ and $m=0$ the period spacing $\Delta P$ between modes of consecutive radial orders as a function of the period $P=2\pi/\omega_{\rm i}$.
In a non-rotating model, the period spacings should be mostly constant with dips related to the chemical stratification near the core \citep{Miglio}.
In rotating stars, the slope of the \emph{g}-mode period spacing pattern is related to the near-core rotation rate \citep{Bouabid, VanReeth16, Ouazzani17}.
As can be seen in Fig.~\ref{fig:l1m0}, the slope is not affected by the magnetic field, but the depth and the spacing of the dips show a significant modification at longer periods (i.e. low frequencies), where the effects of a magnetic field are stronger, as expected from Eq.~\eqref{eq:scaling}.
In addition, the magnetic field induces the presence of peaks close to the dips, thus forming a sawtooth  pattern.
 A similar behaviour is observed for $\ell=2$.
Figure~\ref{fig:l1m0} also illustrates the fact the magnetic shifts scale with the square of the field strength.

According to \citet{VanReeth18}, the effect of differential rotation on gravito-inertial modes is similar to a modification of the Brunt-V\"ais\"al\"a profile, which creates additional dips in the period spacings.
Therefore, magnetic signatures are different from  the signatures of radial differential rotation or stratification.
This is a key property that may help to constrain the structure of internal stellar magnetic fields.

\subsection{Prograde modes}
\label{sec:pro}

\begin{figure}
    \resizebox{\hsize}{!}{\includegraphics{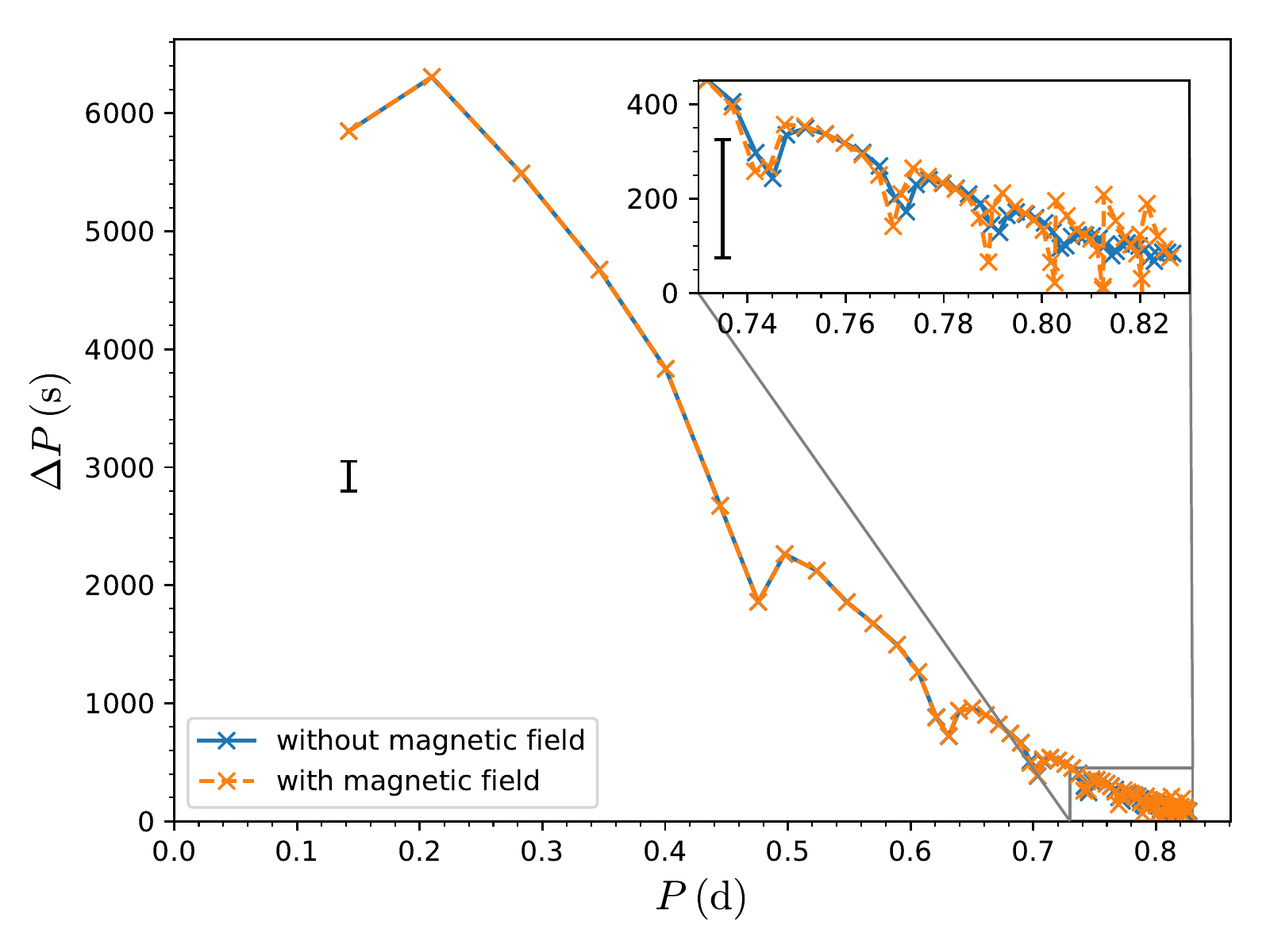}}
    \caption{Same as Fig.~\ref{fig:l1m0}, but for $\ell=1$, $m=1$, and $B_0=1.5\cdot10^5\,{\rm G}$.}
    \label{fig:l1m1}
\end{figure}

Figure~\ref{fig:l1m1} represents the \emph{g}-mode period spacing of modes with $\ell=1$ and $m=1$ and shows that low-frequency modes are strongly affected by the magnetic field, similarly to zonal modes.
The main difference is that period spacings of prograde modes tend to zero for high radial orders while their period remains finite, which makes them potentially harder to distinguish from each other.

\subsection{Retrograde modes}
\label{sec:retro}

\begin{figure}
    \resizebox{\hsize}{!}{\includegraphics{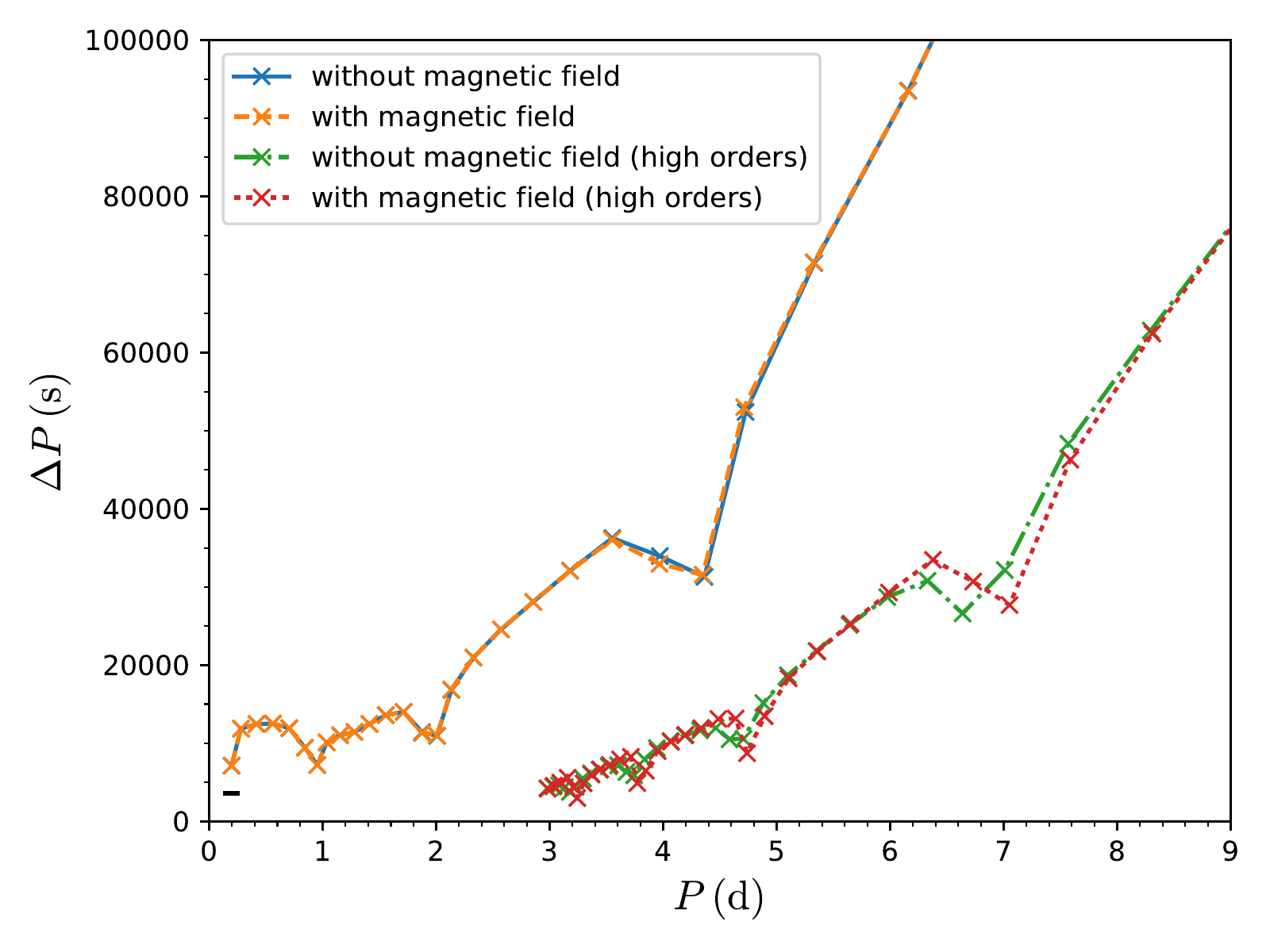}}
    \caption{Period spacings of \emph{g} modes with radial orders from -1 (bottom left) to -26 (top right) and -44 (right) to -74 (bottom) as a function of the period in the inertial frame for $\ell=1$, $m=-1$, and $B_0=1.5\cdot10^5\,{\rm G}$.
    We note that modes from $n=-1$ to $n=-11$ were observed for \hd.
The vertical black bar (very small here) represents a typical observational error bar of 250\,s.}
    \label{fig:l1m-1}
\end{figure}

For retrograde modes, Eq.~\eqref{eq:doppler} may lead to negative frequencies.
Observationally, negative frequencies cannot be distinguished from positive ones.
Therefore, retrograde modes are split into two series of modes.
The corresponding period spacings are plotted in Fig.~\ref{fig:l1m-1}.
Low-radial-order modes (which have short periods in the corotating frame) are only slightly affected by the magnetic field.
In contrast, high-radial-order modes (which have long periods in the corotating frame, but moderate ones in the inertial frame) show clear signatures of the magnetic field.

\subsection{Effect of rotation}
\label{sec:rot}

\begin{figure}
    \resizebox{\hsize}{!}{\includegraphics{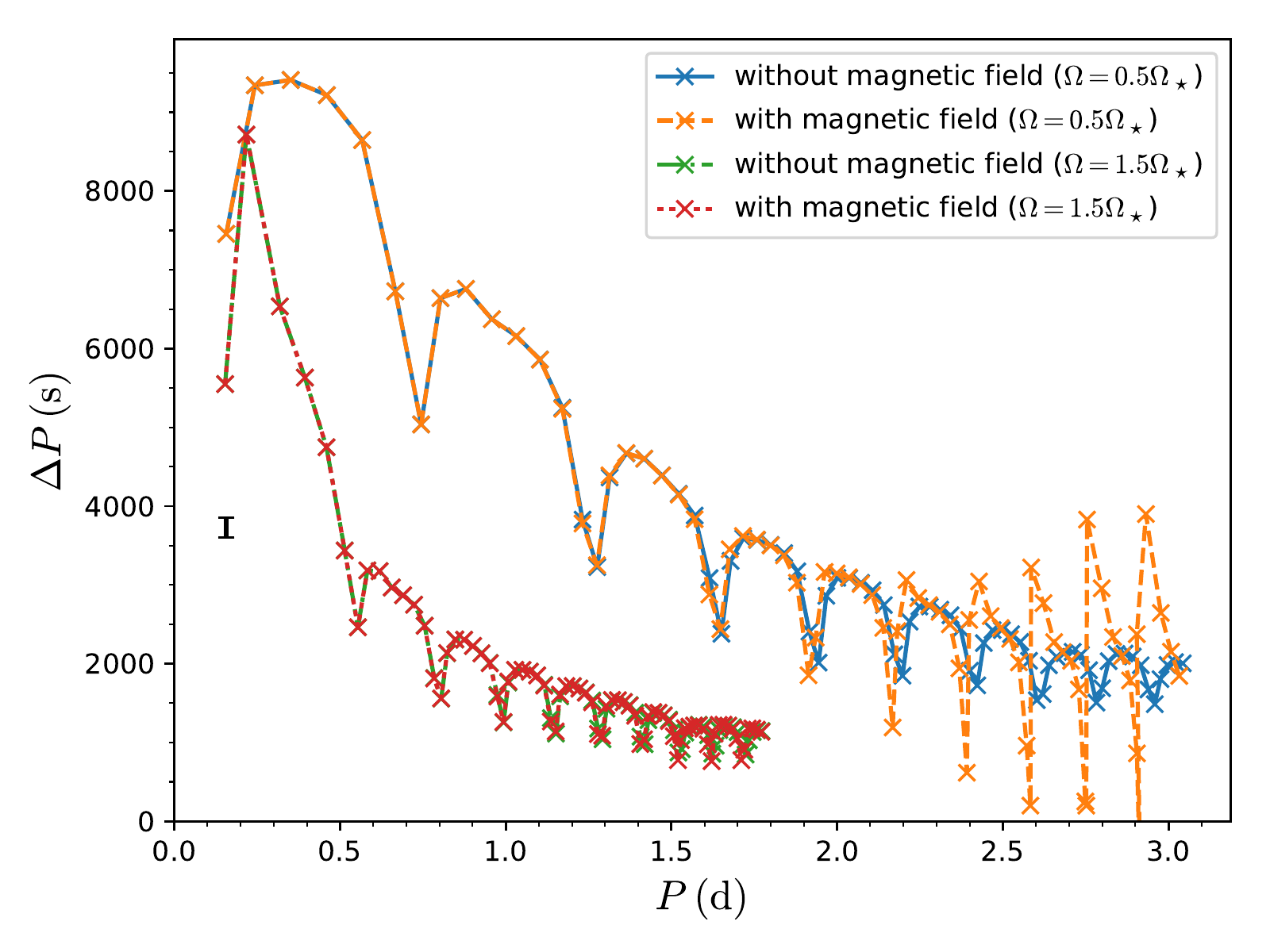}}
    \caption{Same as Fig.~\ref{fig:l1m0}, for $B_0=1.5\cdot10^5\,{\rm G}$ and two different rotation rates: $\Omega=0.5\Omega_{\star}$ and $\Omega=1.5\Omega_{\star}$ (around 30\% and 90\% of the critical rotation rate, respectively).}
    \label{fig:l1m0_rot}
\end{figure}

In this section, we present the computed modes and the associated magnetic frequency shifts with 50\% and 150\% of the measured rotation rate of \hd (around 30\% and 90\% of the critical rotation rate, respectively).
The obtained period spacings are plotted in Fig.~\ref{fig:l1m0_rot}.
It is clear from this figure that increasing the rotation rate significantly decreases the amplitude of the expected magnetic signatures.
This can be  easily explained:  when the Brunt-V\"ais\"al\"a frequency is much higher than the Coriolis frequency $2\Omega$, which is usually  the case, the lower bound for the frequency of gravito-inertial waves $\omega_-$ is close to $2\Omega\cos\theta$ \citep[see e.g.][]{PratLB}.
The Alfv\'en frequency is usually lower than $\omega_-$.
When increasing the rotation rate, $\omega_-$ also increases and moves further away from the Alfv\'en frequency.
Hence, gravito-inertial waves become less sensitive to the presence of the magnetic field with increasing rotation velocity.
This is also consistent with the criterion derived in Eq.~\eqref{eq:criterion}.
It is therefore best to search for the signatures of a magnetic field in period spacing patterns of slow rotators.

For the slower rotation rate considered here, Fig.~\ref{fig:l1m0_rot} displays a negative period spacing value, with a higher-radial-order mode having a shorter period than a lower-radial-order mode.
In reality, this could lead to an avoided crossing between consecutive modes \citep[e.g.][]{Lignieres06}, which would require a non-perturbative treatment of the magnetic field.
This would also make the detection of period spacing patterns significantly more difficult.

\section{Discussion}
\label{sec:conc}

In the present work we investigated the effect of a mixed, axisymmetric, internal large-scale magnetic field, which presumably (but not necessarily) is of fossil origin, on the oscillation frequencies of gravito-inertial modes in the traditional approximation of rotation.
The numerical application to a model of a SPB star showed that high-radial-order modes exhibit significantly distinct \emph{g}-mode period spacing patterns because of the magnetic field compared to non-magnetic stars.
In particular, the magnetic field reveals itself by a sawtooth  pattern in the period spacing morphology, rather than the typical dips that occur due to mode trapping caused by a $\mu$-gradient left behind by the shrinking convective core \citep{Miglio}.
Thus, the search for such predicted patterns is a way to discover internal magnetic fields from gravity modes.
In addition, computations at different rotation rates highlighted that magnetic signatures decrease with rotation.
As a consequence, it is crucial to take rotation into account when computing magnetic frequency shifts.
Otherwise, the strength of the magnetic field needed to explain observed signatures  would be drastically underestimated.
It is noteworthy that the two \emph{Kepler} SPB stars that were modelled in detail are both slow to moderate rotators.
Their modelling revealed shortcomings in their frequency fits for the modes with higher periods (hence higher orders) \citep{Moravveji15, Moravveji16}.
Our work provides a good reason to revisit the modelling of these stars, assuming the presence of a magnetic field.

The predicted magnetic signatures seem to be  related (at least partly) to the dips in the period spacings created by chemical gradients.
These gradients are a consequence of stellar evolution and are thus not present in stars on the zero-age main sequence (ZAMS).
Recently, \citet{Mombarg} determined the asteroseismic ages of 37 $\gamma$~Dor stars, finding several of those to be near the ZAMS, even though they show significant dips in their \emph{g}-mode period spacing patterns.
Therefore, it would be interesting to determine whether the signatures of a large-scale magnetic field are measurable for such stars.
More generally, we intend to perform a systematic exploration of magnetic effects as a function of stellar parameters, such as mass, core mass, metallicity, and magnetic field strength in the near-core region, where \emph{g} modes are most sensitive.

In order to verify that the TAR is valid in the considered regime and that the magnetic field is weak enough to use the perturbation theory, it would be interesting to compare the obtained frequencies to two-dimensional computations of modes, for example with the Two-dimensional Oscillation Program \citep[TOP;][]{Reese06} or the Adiabatic Code of Oscillation including Rotation \citep[ACOR;][]{Ouazzani12}.
However, neither of these oscillation codes accounts for the magnetic field at the moment.
Another application of these codes would be to apply the perturbative theory to gravito-inertial modes computed in centrifugally deformed stars.

In the case of \hd, magnetic signatures could not be extracted because too few mode frequencies were identified, especially for high-radial-order modes, where the effect of the magnetic field is strongest.
We note that this star was observed with CoRoT.
In principle, stars observed with \emph{Kepler} should allow  many more modes to be detected.
Another possible reason for the low number of identified frequencies is that only a fraction of computed modes may actually be excited.
In addition, some modes may not be visible because of surface cancellation effects and the inclination of the star.
However, some stars do exhibit gravity modes with radial orders down to -50 (see \citealt{VanReeth15b} for F stars and \citealt{Papics17} for B stars).
When a series of these modes is observed, our results show that magnetic signatures could be extracted for a field of a reasonable strength.
 \citet{VanReeth15b} give estimates of the error on the period spacings for $\gamma$~Dor stars that range from roughly 20 to 1000\,s (an average of 250\,s was used in Figs.~\ref{fig:l1m0}--\ref{fig:l1m0_rot}), which is in some cases smaller than the expected magnetic signatures.

Although our mixed (poloidal and toroidal) magnetic configuration is a significant improvement compared to previous studies, it is still  relatively simple (confined axisymmetric dipole).
First, a more realistic work would require  considering a magnetic field that is not confined within the star.
The formalism derived in this work is  still valid {a priori}, but since the density becomes very small at the surface, surface effects are to be expected \citep[see e.g.][]{Bigot}.
Modelling these effects would require a non-perturbative treatment of the magnetic field \citep[e.g.][]{Morsink}.
A similar treatment would also be needed to investigate the effect of a very strong internal magnetic field.
Second, many stars, including \hd, have a dipolar magnetic field inclined with respect to the rotation axis.
Our formalism needs to be generalised to the case of a non-axisymmetric field to predict the influence of the obliquity angle on the magnetic signatures, but this is beyond the scope of the present paper and is the subject of future work.
Finally, general three-dimensional non-axisymmetric configurations should be studied.
They will be of natural interest for fields generated by dynamos \citep[e.g.][and references therein]{Brun05, Augustson16, BrunBrowning}, complex fossil fields \citep{Donati,Braithwaite08,Kochukhov}, and unstable fields \citep[e.g.][]{Braithwaite06,Braithwaite07,Brun07,Zahn07}.
Seismic magnetic signatures could also be used to help detect and constrain magnetic fields of Vega-like stars \citep{Lignieres09}.

Another possible follow-up of this work is to generalise it to the case of differential rotation.
A full treatment of differential rotation would probably make the present formalism unusable.
However, non-axisymmetric magnetic fields are known to inhibit differential rotation \citep{Moss,Spruit}.
Thus, a perturbative treatment of differential rotation might be sufficient.

\begin{acknowledgements}
    The research leading to these results received funding from the European Research Council (ERC) under the European Union's Horizon 2020 research and innovation program (grant agreements No.~647383: SPIRE with PI S.M. and No.~670519: MAMSIE with PI C.A.).
    V.P. and S.M. acknowledge support from the CNES PLATO grant at CEA/DAp.
    V.P. acknowledges the International Space Science Institute (ISSI) for supporting the SoFAR international team\footnote{\url{http://www.issi.unibe.ch/teams/sofar/}}.
    The authors thank the referee for the useful comments.
\end{acknowledgements}

\bibliographystyle{aa}
\bibliography{refs}

\appendix
\onecolumn

\section{Hough functions}
\label{sec:hough}

The radial Hough function \citep{Hough} is defined by $H_r(\theta) = f(\mu=\cos\theta)$, where $f$ is the solution of the eigenvalue equation
\begin{equation}
    \frac{1-\mu^2}{1-\nu^2\mu^2}\frac{{\rm d}^2f}{{\rm d}\mu^2} - \frac{2\mu(1-\nu^2)}{(1-\nu^2\mu^2)^2}\frac{{\rm d}f}{{\rm d}\mu}+\left[\frac{m\nu(1+\nu^2\mu^2)}{(1-\nu^2\mu^2)^2}-\frac{m^2}{(1-\mu^2)(1-\nu^2\mu^2)}\right]f = \lambda f,
\end{equation}
and $\nu=2\Omega/\omega$ is the spin factor.
In the non-rotating case, the eigenvalue $\lambda$ reduces to $-\ell(\ell+1)$, where $\ell$ is the angular degree of the mode, and $H_r$ reduces to the classical associated Legendre polynomial $P_\ell^m$.

The latitudinal and azimuthal Hough functions are respectively given by
\begin{equation}
    H_\theta \sin\theta = \frac{H_r' \sin\theta - m\nu H_r\cos\theta}{1-\nu^2\cos^2\theta}
\end{equation}
and
\begin{equation}
    H_\varphi \sin\theta = \frac{mH_r-\nu H_r'\sin\theta\cos\theta}{1-\nu^2\cos^2\theta},
\end{equation}
where $'$ denotes here a total latitudinal derivative.
It is obvious from these definitions that $H_\varphi$ has the same parity as $H_r$ with respect to $\theta=\pi/2$, whereas $H_\theta$ has the opposite parity.

In the present work, we computed Hough functions using the implementation based on Chebyshev polynomials proposed by \citet{Wang}.

\section{Non-zero-average terms of the Lorentz work}
\label{sec:terms}

In this section, we list all the non-zero-average terms of the work of the Lorentz force $\delta\vec F_{\rm L}\cdot\vec\xi^*$ (for convenience, we take here the product with $\mu_0$).
Interestingly, all these terms involve either only poloidal components of the magnetic field, or only the toroidal component, and we group them accordingly.
In addition, all terms have a purely radial part multiplied by a purely latitudinal part, and the prime symbol ($'$)   denotes a total derivative, either radial or latitudinal depending on the context.

\subsection{Poloidal terms}

Defining $A = [(rb_\theta)'+b_r]$, the terms involving poloidal components are
\begin{align*}
    &-m\frac{\xi_{\rm h}b_\theta A\xi_r^*}{r^2}H_rH_\varphi\sin\theta
    +\frac{(r\xi_rb_\theta)'A\xi_r^*}{r^2}H_r^2\sin^2\theta
    -\frac{(r\xi_{\rm h}b_r)'A\xi_r^*}{r^2}H_rH_\theta\sin\theta\cos\theta
    +\frac{\xi_rb_\theta A\xi_{\rm h}^*}{r^2}H_\theta(H_r\sin^2\theta)'
    -\frac{Ab_r|\xi_{\rm h}|^2}{r^2}H_\theta(H_\theta\sin\theta\cos\theta)'\\
    &+m\frac{Ab_r|\xi_{\rm h}|^2}{r^2}H_\theta H_\varphi\cos\theta
    -m\frac{(\xi_{\rm h}b_\theta)'b_\theta\xi_r^*}{r}H_rH_\varphi\sin\theta
    +\frac{(r\xi_rb_\theta)''b_\theta\xi_r^*}{r}H_r^2\sin^2\theta
    -\frac{(r\xi_hb_r)''b_\theta\xi_r^*}{r}H_rH_\theta\sin\theta\cos\theta\\
    &+\frac{b_\theta^2|\xi_r|^2}{r^2}H_r\sin\theta\left[\frac{(H_r\sin^2\theta)'}{\sin\theta}\right]'
    -\frac{\xi_{\rm h}b_rb_\theta\xi_r^*}{r^2}H_r\sin\theta\left[\frac{(H_\theta\sin\theta\cos\theta)'}{\sin\theta}\right]'
    +m\frac{\xi_{\rm h}b_rb_\theta\xi_r^*}{r^2}H_r\sin\theta\left(H_\varphi\frac{\cos\theta}{\sin\theta}\right)'
    +m\frac{(\xi_{\rm h}b_\theta)'b_r\xi_{\rm h}^*}{r}H_\theta H_\varphi\cos\theta\\
    &-\frac{(r\xi_rb_\theta)''b_r\xi_{\rm h}^*}{r}H_rH_\theta\sin\theta\cos\theta
    +\frac{(r\xi_{\rm h}b_r)''b_r\xi_{\rm h}^*}{r}H_\theta^2\cos^2\theta
    -\frac{\xi_rb_rb_\theta\xi_{\rm h}^*}{r^2}H_\theta\cos\theta\left[\frac{(H_r\sin^2\theta)'}{\sin\theta}\right]'
    +\frac{b_r^2|\xi_{\rm h}|^2}{r^2}H_\theta\cos\theta\left[\frac{(H_\theta\sin\theta\cos\theta)'}{\sin\theta}\right]'\\
    &-m\frac{b_r^2|\xi_{\rm h}|^2}{r^2}H_\theta\cos\theta\left(H_\varphi\frac{\cos\theta}{\sin\theta}\right)'
    +\frac{(r\xi_{\rm h}b_r)'b_\theta\xi_{\rm h}^*}{r^2}H_\varphi(H_\varphi\sin\theta\cos\theta)'
    +\frac{b_\theta^2|\xi_{\rm h}|^2}{r^2}H_\varphi[\sin\theta(H_\varphi\sin\theta)']'
    -m^2\frac{b_\theta^2|\xi_{\rm h}|^2}{r^2}H_\varphi^2\\
    &+m\frac{(r\xi_rb_\theta)'b_\theta\xi_{\rm h}^*}{r^2}H_rH_\varphi\sin\theta
    -m\frac{(r\xi_{\rm h}b_r)'b_\theta\xi_{\rm h}^*}{r^2}H_\theta H_\varphi\cos\theta
    -m\frac{\xi_rb_\theta b_r\xi_{\rm h}^*}{r^2}H_\varphi\frac{\cos\theta}{\sin^2\theta}(H_r\sin^2\theta)'
    +m\frac{b_r^2|\xi_{\rm h}|^2}{r^2}H_\varphi\frac{\cos\theta}{\sin^2\theta}(H_\theta\sin\theta\cos\theta)'\\
    &-m^2\frac{b_r^2|\xi_{\rm h}|^2}{r^2}H_\varphi^2\frac{\cos^2\theta}{\sin^2\theta}
    +\frac{(r\xi_{\rm h}b_r)''b_r\xi_{\rm h}^*}{r}H_\varphi^2\cos^2\theta
    +\frac{(\xi_{\rm h}b_\theta)'b_r\xi_{\rm h}^*}{r}H_\varphi\cos\theta(H_\varphi\sin\theta)'.
\end{align*}

\subsection{Toroidal terms}
\label{sec:tor}

The terms that involve the toroidal component are
\begin{align*}
    &2\frac{(r\xi_rb_\varphi)'b_\varphi\xi_{\rm h}^*}{r^2}H_rH_\theta\sin\theta\cos\theta
    +2\frac{b_\varphi^2|\xi_{\rm h}|^2}{r^2}H_\theta\cos\theta(H_\theta\sin\theta)'
    +2m\frac{b_\varphi^2|\xi_{\rm h}|^2}{r^2}H_\theta H_\varphi\cos\theta
    +\frac{(r\xi_rb_\varphi)''b_\varphi\xi_r^*}{r}H_r^2\sin^2\theta\\
    &+\frac{(\xi_{\rm h}b_\varphi)'b_\varphi\xi_r^*}{r}H_r\sin\theta(H_\theta\sin\theta)'
    +\frac{(r\xi_rb_\varphi)'b_\varphi\xi_{\rm h}^*}{r^2}H_\theta(H_r\sin^2\theta)'
    +\frac{b_\varphi^2|\xi_{\rm h}|^2}{r^2}H_\theta[\sin\theta(H_\theta\sin\theta)']'
    -m^2\frac{b_\varphi^2|\xi_{\rm h}|^2}{r^2}H_\theta^2\\
    &+\frac{(r\xi_rb_\varphi)'(rb_\varphi)'\xi_r^*}{r^2}H_r^2\sin^2\theta
    +\frac{\xi_{\rm h}b_\varphi(rb_\varphi)'\xi_r^*}{r^2}H_r\sin\theta(H_\theta\sin\theta)'
    +m\frac{\xi_rb_\varphi(rb_\varphi)'\xi_{\rm h}^*}{r^2}H_rH_\varphi\sin\theta
    -m^2\frac{b_\varphi^2|\xi_r|^2}{r^2}H_r^2.
\end{align*}
The first eight terms in the previous equation are those predicted to have a significant impact on frequency shifts for a strong toroidal field.

\end{document}